%% file: Selfish.tex
\documentclass[a4paper,10pt]{article}
\usepackage{graphicx}

\begin{document}
\pagestyle{plain}
\setcounter{page}{0}

\parindent 8mm
\pagestyle{empty}

\vspace*{2cm}
\title{The Evolution of Cellar Automaton based on Dilemmma Games with Selfish Strategy}
\author{Norihito Toyota}
\date{Hokkaido Information University,
59-2 Nishinopporo Ebetsu City, Japan \\
E-mail:toyota@do-johodai.ac.jp
}

\maketitle
\baselineskip 5mm
\pagestyle{empty}
\begin{abstract}
We have proposed two new evolutionary rules on spatio-iterated games that is not mimic evolution of strategies, and mainly discussed the Prisoner's Dilemma game \cite{toyota2} by the two evoutionary rules \cite{toyota3}. 
In this paper we focus the first rule, that is, the selfish evolutionary rule for various dilemma games. 
In contrast to the Prisoner's Dilemma, there are gererally rich pase structures in the dilemma games. 
First we analytically clear the structure to present phase diagrams in the various dilemma games.    
Forthermore we simulate the time evolution of the soatio-games in the some representatives of the parameters according to the phase diagrams.    
  Including some mutations, detail investigations are made by a computer simulation  for five kinds of initial configurations.  
As results we find some dualities and game invariant properties. 
They show a sort of bifurcation as a mutation parameter are varied. 
In the path from one period to two one some common features are observed in most of games and some chaotic behaviors appear in the middle of the transition.  
Lastly we estimate the total hamiltonian, which is defined by the sum of the total payoff of all agents in the system, and show that the chaotic period is best from the perspective of the payoff.  
We also made some primitive discussions on them.

\end{abstract}
\bf{key words:} 
\it{Chicken Game,Prisoner's Dilemma, Stug Hunt Game, Cellar Automaton, Spatio-Evolutionary Game,  Phase Structure }

\rm
\normalsize
\baselineskip 6mm
\pagestyle{empty}
\thispagestyle{empty}

\renewcommand{\thefootnote}{\fnsymbol{footnote}} 
\rm
\normalsize
\baselineskip 6mm

\section{Introduction}
Traditionally differential equations have been applied to understand various phenomena in the nature and the artificial, especially physical phenomena, since they were used to construct the classical dynamics by I. Newton. 
It, however,  seems that some phenomena are rather relevant to be studied as they follow descrete dynamics.
 They may include biological, social, ecological phenomena and so on, which have a close relationship to "information".
 The game theory among many aproaches give one of interesting view of them. 
In this paper, we discuss various games with Dilemma, some kinds of the Chicken games,  Prisoner's Dilemma (PD), the Stag Hunt game  which have been pointed out to be dilemma games by Poundstone \cite{Poun},\cite{Rapo}.  

They are symmetric games played by two agents and the payoff function acquired 
by taking each action is discribed by a bimatrix form. 
A set of the Chicken games, which include the so-colled the Chicken game, th Hero game, the Leader game, are dilemma games because they have two Nash equilibria. 
In this paper we use the term  "the Chicken game"   in a narrow sense. 
In the meanwhile,
 PD faces dilemma in the sense that   $pareto \; optimal \neq Nash\; equilibrium$\cite{Okad}(the  deadlock game does not face any dilemma \cite{Poun}).

Especially many researches have been made to resolve the dilemma, $pareto \; optimal \neq Nash\; equilibrium$, and how can we obtain cooperation rationally in PD.   
Based on PD, iterated games have been studied and Axelrod \cite{Axel} has shown that cooperation can emerge as a norm in a society comprised of individuals with selfish motives. 
Moreover spatial version of PD has been discussed so that a defection is to be only evolutionary stable strategy (ESS) \cite{Smit} if each agent interacts with any other agents. This aspect drastically changes if a spatial structure of the population is considered. If the interaction between agents is locally restricted to their neighbors \cite{Nowa},\cite{Nowak}, a stable coexistence between cooperators and defectors become possible under certain conditions \cite{Nowa},\cite{Nowak}. 
In the case, one agent plays a game with their neighbors and in next step the agent take the same strategy as that of the agent that acquired hightest payoff among the neighbors,  which reflects Dawin's theory, sometimes including "mutation"\cite{Fort}. 
Then it is assumed that all agents play at same time and follow the same way. 
Recently the evolution of the spatio-structured PD has been systematically explored in details in Ref.\cite{Schw}.

We have discussed two other evolutionary rules within the framework of the spatio-structured games where agents interact locally on the neighborhood cells\cite{toyota2},\cite{toyota3},\cite{toyota4}.
One way to realize it is to change his(her) action to new one when the agent would get larger payoff if he(she) had chosen the opposite action. 
This  is called the $selfish\;\; rule$ in this paper. 
Second one is to change the action like totalitarian, that is, if the total payoff yielded in the whole system increases when an agent changes the strategy. 
We call this evolutionary rule the $totalitarian\;\; rule$. 
This evolutionary role is expected to lead to full cooperative action but it do not so\cite{toyota2},\cite{toyota3}. 

In this paper we only discuss the former case in various type dilemma games and 
argue general properties lurking in the time evolution of them. 
Then we introduce the rule to a kind of mutation based on the Gibbs distribution. 
As a result of the simulations, we find some general properties when one periodic motion of the population with cooperation undergos a transition to two periodic one, regardless of changing  payoff parameters. 
 In the  bifurcation, some chaotic behabiours generally appears. 
Some considerations will be given for it.    

 After the Introduction, we discuss our evolutionary rule and give the phase structure of games discussed here based on an analytical study in the section 2. In the section 3 simulation results for a set of Chicken games are obtained based on the results in the section 2. 
There the time serie of the population of each action is studied in details and some analytical arguments are also given. The  section 4 is devoted to the analysis of other dilemma games. 
Concluding remarks are given in the final section 5. 

\section{Phase Structure of Dilemma Games}

\subsection{Spatial structured game as cellar automaton}
Two type agents, cooperators C and defectors D are considered on cells with the size $N=n\times n$. 
The distinction is made by means of the suffixes $C$ and $D$. 
The total number of agents are given by 
\begin{equation}
N=N_C + N_D,
\end{equation}
where $N_C$ and $N_D$ are the population number of C-agents and D-agents, respectively. 
The spatial distribution of agents are considered as a two dimensional cellar automaton (CA) consisting of $N$ cells, where each cell is identified by the index $i \in N$ refering its spatial position and the state C or D sits on the cell. The state space of all possible configurations is of order $2^N$.  We asumme that  an agent $i$ simultaneously plays with 4 neighbors of $i$ with a same strategy C or D, and so the game essentially reduces to 2-person game. All agents in the system asynchronously play a geme every round. Then the game can be described by a peyoff bimatrix such as Table 1.

\subsection{Formulation of dilemma games}
We give some general and analytic discussions on symmetric games played by two agents and give the phase diagrams of them.
The symmetric games with two choices played by two agents are generally defined by a payoff table which is often represented a bimatrix such as Table 1, where two actions C and D show Cooperation and Defection, respectively. 
$i$ and $j$ distinguishes two agents. 

\begin{table}[h]\centering
\caption{General form of a payoff table of 2$\times$ 2 symmetric game.
 The left and right variables in the parenthesis show 
the payoffs of the agent $i$ and $j$, respectively. }
\begin{tabular}{|c|c|c|} \hline
  &\makebox[15mm]{C$_j$} & \makebox[15mm]{D$_{j}$} \\ \hline
C$_{i}$ & ($R,R$) & ($S,T$) \\ \hline
D$_{i}$ & ($T,S$) & ($P,P$) \\ \hline
\end{tabular}
\end{table}
 Each game is classified  according to the magnitude of $R$, $S$, $T$ and $P$. 
The list of games explored in this paper is the following. \\
(1)Chicken (exploiter)game for $T>R>S>P$.\\
(2)Hero game for $ S>T>R>P$.\\
(3)Leader game (which essentially includes the battle of the sexes) for $T>S>R>P$.\\
(4)PD for $T>R>P>S$.\\
(5)Stag Hunt game for $R>T>P>S$.\\  

(1), (2) and (3) are included within a set of the Chicken games in a broad sense, and they and (5) have two Nash equilibria. 

We consider a sort of 2 dimensional cellar automaton. 
One agent exits at a cell on torus (i.e. we consider a periodic boundary condition) with $n\times n=N$ cells. 
A agent play a game with the four agents stood in the Neumann neighborhood with a fixed action C or D  at each step. 
The payoff acquired in each step is estimated by using Table. 1.  
Under this situation, if an agent tried a different choice, the agent might get more payoff from the Neumann neighborhood.  
If so, the agent changes to the opposite action at the next step in the  selfish rule. 

Let's estimate the increment of a payoff, $\Delta P (C\rightarrow D)$ , when an agent change the action from C to D. It generally depends on the population of C agents of his(her) Neumann neighborhood. They are given as follows;
\begin{eqnarray}
\Delta P (C\rightarrow D)[0,4]&=&4(P-S), \\
\Delta P (C\rightarrow D)[1,3]&=&(T-R)+3(P-S), \\
\Delta P (C\rightarrow D)[2,2]&=&T-R+P-S, \\
\Delta P (C\rightarrow D)[3,1]&=&3(T-R)+(P-S), \\
\Delta P (C\rightarrow D)[4,0]&=&4(T-R),
\end{eqnarray}
where the numbers of the inside of the square brackets the represent the populations of C and D in the Neumann neighborhood, respectively. 
 That is, $\Delta P (C\rightarrow D)[k,\ell]$ means the increment of the payoff obtained by an agent when the target agent is  surrounded by $k$ cooperators and $\ell$ defectors.   
When an agent change the action from D to C, the signs in the Eqs. (2)-(6) reverse. 
While the order of the magnitude of $R$, $S$, $T$ and $P$ is only essential in usal game theories, the values of $R$, $S$, $T$ and $P$ themselves play an important role in an iterated game. 
By changing the four values, the signs in above equations may go into reverse. 
To vary all of them  independently is really  too complex  to analyze them. Instead of varying all of them, we change only largest parameter and lowest parameter in each game. 
For example of the Chicken game with $T>R>S>P$, we change only $T$ and $P$, and fix $R$ and $S$ to some correct values, because $T$ can take from $R$ to $\infty$ and $P$  from $S$ to $-\infty$. $R$ and $S$ are restricted by $T$ and $P$, which  they can move so freely. 
In this paper we basically follow values given by Okada\cite{Okad} for restricted parameters. 

\subsection{Phase structure of dilemma games}
We investigate the phase structure of each game based on the knowledge given above. \\

(1)Chicken Game\\
We adopt $R=5$ and $S=-4$ (according to Okada \cite{Okad} and variables are $T$ and $P$. Then 
\begin{equation}
\Delta P (C\rightarrow D)[0,4]=4(P-S) <0 \;\;\mbox{and}\;\;
\Delta P (C\rightarrow D)[4,0]=4(T-R)>0,
\end{equation}
where the inequality are due to the condition of $T>R>S>P$ of the Chicken Game. In order to explore other equations, we observe the points where the values of the equations are zero;
\begin{eqnarray}
\Delta P (C\rightarrow D)[1,3]&=&(T-R)+3(P-S)=0, \\
\Delta P (C\rightarrow D)[2,2]&=&T-R+P-S=0, \\
\Delta P (C\rightarrow D)[3,1]&=&3(T-R)+(P-S)=0. 
\end{eqnarray}
The different evolutionary behaviour is expected to appear in each region  
divided by the boundaries given by Eqs. (8)-(10) on $T-P$ space. 
They are summarized in Fig.1. \\

\input{ChikinH2.tex}\\

\vspace*{5mm}

The part with  oblique lines in each Fig. is only available for the corresponding game. 
The figures in parentheses on real lines in Figs. correspond to the nunmers of equations.\\

(2)Hero game\\
We adopt $T=6$ and $R=5$ and variables are $P$ and $S$. Then 
equation (7) also holds. 
For other equations, we obtain in a similar way to the Chicken game;
\begin{eqnarray}
\Delta P (C\rightarrow D)[1,3]&=&a+3(P-S)=0, \\
\Delta P (C\rightarrow D)[2,2]&=&a+(P-S)=0, \\
\Delta P (C\rightarrow D)[3,1]&=&3a+(P-S)=0, 
\end{eqnarray}
where we generally introduce $a \equiv (T-R) $ and in this case $a=1$.  
Fig.2 shows the phase structure of Hero game. \\

(3)Leader game\\
Eq. (7) also hols, which is a common prperty in a set of the Chicken games. 
 For other equations with $R=-4$ and $S=5$, we obtain, 
\begin{eqnarray}
\Delta P (C\rightarrow D)[1,3]&=&T+3P -11=0, \\
\Delta P (C\rightarrow D)[2,2]&=&T+P-1=0, \\
\Delta P (C\rightarrow D)[3,1]&=&3T + P +7=0, 
\end{eqnarray} 
Fig.3 shows the phase structure of the Leader game. \\

(4)PD game\\
In PD, the payoffs have to satisfy  the following condition;
$T>R>P>S$, and so Eqs. (2)-(6) are trivially positive. 
So any sigunificant structure can be found. 
We here make a brief comment on PD. 
For iteration of PD, an additional condition is usually imposed:
\begin{equation}
2 R> S+T.
\end{equation}
The second condition, however, is considered to have not any essential meaning for the structured spatio-game. 
That the breakdown of this condition (17) really induces some interesting phenomena in spatio PD with totalitarian rule has been pointed out \cite{toyota2},\cite{toyota3}. \\

\input{LeaderSt6-3.tex}\\

(5)Stag Hunt game\\
Eqs. (2) and (6)  are positive and negative in this game, respectively. 
For other equations, we choose $P=-3$ and $T=5$ and obtain
\begin{eqnarray}
\Delta P (C\rightarrow D)[1,3]&=&12-3R-S=0, \\
\Delta P (C\rightarrow D)[2,2]&=&2(2-R-S)=0, \\
\Delta P (C\rightarrow D)[3,1]&=&-4-R-3S=0.  
\end{eqnarray} 
We give the phase structure of this game in Fig.4. \\





\subsection{Selfish evolution}
For an iterated game, we assume the selfish evolutionary rule in this paper. 
The selfish evolution is to change present action to opposite one at next step 
only in  the situation that the (target) agent would get lager payoff if he(she) took  the opposite action, independent of other agents. Suppose that all the agents follow this rule and each agent updates its rule in regular order on cells, asynchronously. 

In the simulation experiment of the next section, we introduce mutant agents with some mutation probability $\mu$ with respect to the Bolzmann-like distribution among regular agents that obey the selfish rule;
\begin{equation}
\mu \sim \exp [-|\Delta P| q],
\end{equation} 
 where $q$ corresponds to a temperature in thermodynamics. 
 Agents do not obey the selfish rule with the probability $\mu$.  
In a simulation we take $\mu$ from $0$ to $\infty$.

\section{Simulation of Chicken like games}
In this section we mainly discuss the Chicken game as a representative in datails. 
The investigations can be made in a similar way for the other games.  
For $T$ and $P$, we choose some representatives from regions in Fig.1 and on the borderlines between two or three regions. 
Really we explore the following nine points
$$(P,T)=(-12,5),(-11,6),(-10,7),(-9,8),(-8,9),-(7,10), (-6,11),(-5,12), (-4,13). $$
These are noted in Fig. 1.

As initial states we explore the following five cases; \\
(a) All states are C.\\
(b) All states but  one cell in the center are C.\\
(c) All states are D.\\
(d) All states but one cell in the center are D.\\
(e)Random state with $N_C : N_D=1:1.$\\
These will uncover outcome of initial state dependence. 
Agents play asynchronously in order on the lattice (a random order will also discussed later) with the size $N=12 \times 12$. 
Though the lattice size  $n=12$ seems to be too small, we have ascertained that essential results are unchanged by magnifying the lattice size except for some cases pointed specially. 
While configurations immediately converge in almost every case,  some interesting cases show rather complex behaviors such as chaotic-like behaviors.

\subsection{Chaos-like behaviour and bifurcation}
First of all, we study the cases (a)-(d) as initial states. 
Any phase transition does not occur contrasted with PD game in totalitarian rule\cite{toyota2}. 
At $q=0$,  in the all cases except for the peculiar points, $(P,T)= (-12,5)$ and  $(P,T)= (-4,13)$ which exist on the border of the Chicken game with others, 
the population of cooerators  converges to $N_C (t)= N/2$ after large step $t$. 
Because of one period, we call the state $P1$. 
This makes a checked pattern consist of C and D. 
On the one hand, at $q \rightarrow \infty$, all cases come to 2 period state $P2$. How obout intermediate $q$ ? 
We note that in all cases common behaviors are observed in the intermediate value of $q$. 
Some chaotic behaviour and intermittent chaos appear. These are showed in Figs.5 and 6.  

At a large $q$, the state converge to $P1$, and as $q$ becomes smaller, $C(t)$ begin fluctuating  and comes to a chaotic state. As $q$ is made  further smaller, the amplitude of $C(t)$ increses and chaotic behaviour (C-state) become more and more conspicuous (Fig. 5). Of course since we now consider a finite system, it can not be actually chaos.   As $q$ is made  further and further smaller, an intermittent like chaos ($CI-state$) appears (Fig. 6).  So the parts with small amplitude begin clustering but the parts with a large amplitude also appears at small $t$. 

\addtocounter{figure}{4}
\begin{figure}[t]
\begin{center}
\includegraphics[scale=1.0]{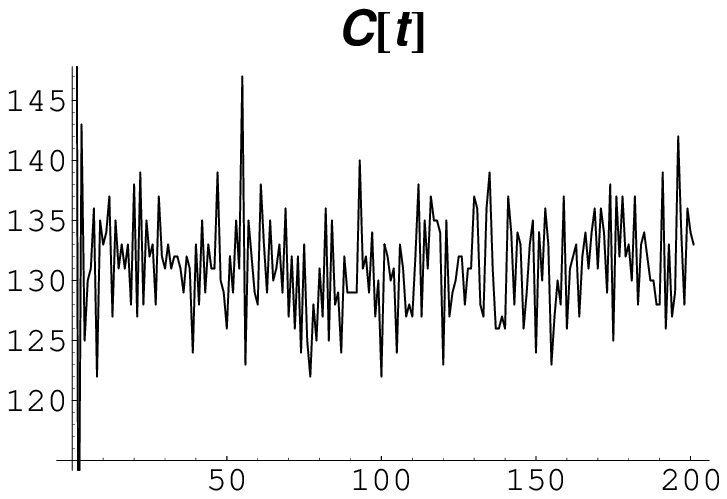}
\caption{Typical chaos-like state in $C(t)$ with $n=14$, $q=0.25$ and $(P,T)=(-11,6)$.}
\end{center}
\end{figure}

\begin{figure}[t]
\begin{center}
\includegraphics[scale=1.0]{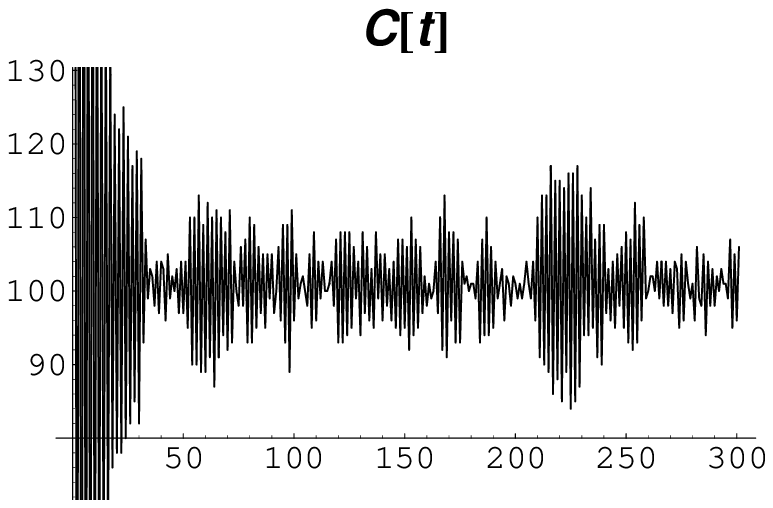}
\caption{Typical $CI$ state in $C(t)$ with $n=14$, $q=0.025$ and $(P,T)=(-11,6)$.}
\end{center}
\end{figure}

After that, the $BI-state$, which has initially $P2$-lile state and releases some chaotic lumps as   $t$ grows larger, appears (see Fig. 7). 
At next stage of smaller $q$, a linerly damped oscillation arises but it does not completely  reduces to one value  but fluctuates around $N_C=N/2$ protractedly($D-state$, Fig. 8). In further smaller $q$, the small fluctuation grows larger ($DP2-state$, Fig. 9) and it spreads over $t$. In the end it converges to the complete $P2$ (Fig. 10). The all-C and the all-D  repeat by turns. 
These are the fundamental pattern in the bifurcation of $P1 \rightarrow P2$. 
They are summarized as follows;
$$P1 \rightarrow C \rightarrow CI \rightarrow DI \rightarrow D \rightarrow P2.$$
\begin{figure}[t]
\begin{center}
\includegraphics[scale=1.0]{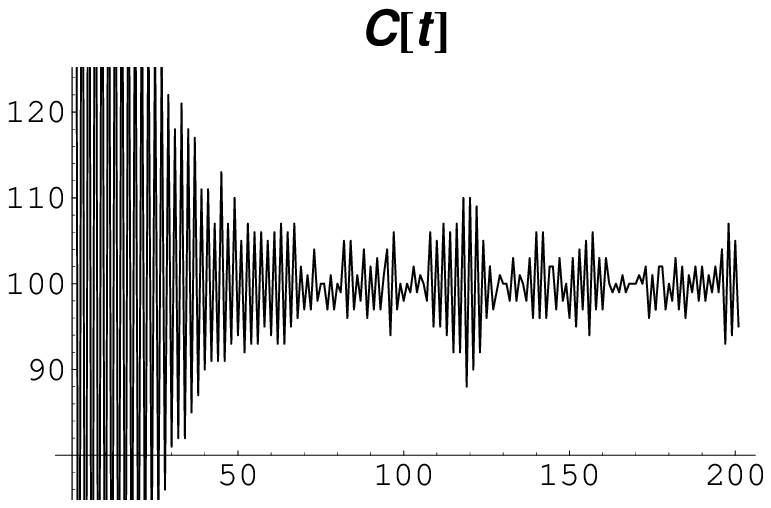}
\caption{Typical DI state in $C(t)$ with $n=14$, $q=0.005$ and $(P,T)=(-10,7)$.}
\end{center}
\end{figure}

\begin{figure}[t]
\begin{center}
\includegraphics[scale=1.0]{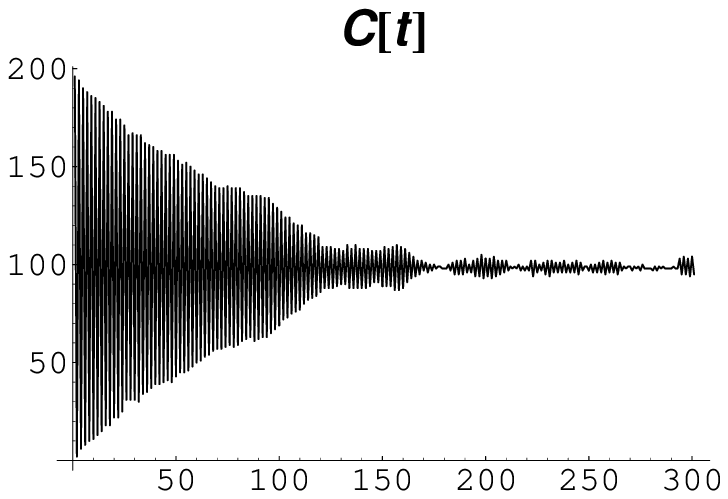}
\caption{Typical $D-$state in $C(t)$ with $n=14$, $q=0.005$ and $(P,T)=(-11,6)$.}
\end{center}
\end{figure}

\begin{figure}[t]
\begin{center}
\includegraphics[scale=1.0]{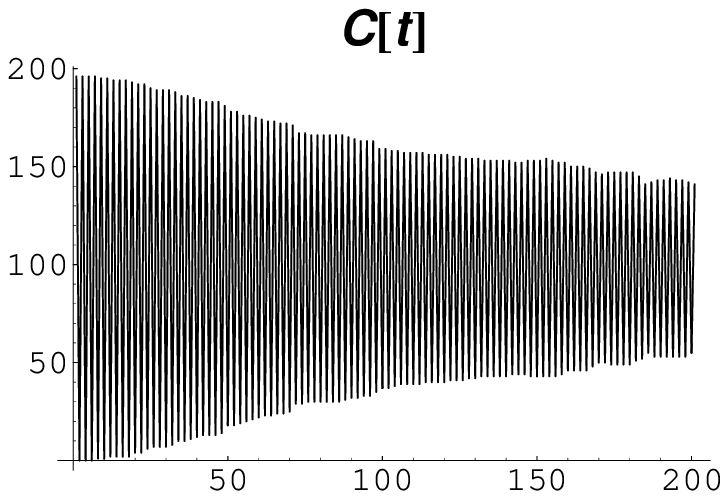}
\caption{Typical $DP2$-state in $C(t)$ with $n=14$, $q=0.0005$ and $(P,T)=(-9,8)$.}
\end{center}
\end{figure}

\begin{figure}[t]
\begin{center}
\includegraphics[scale=1.0]{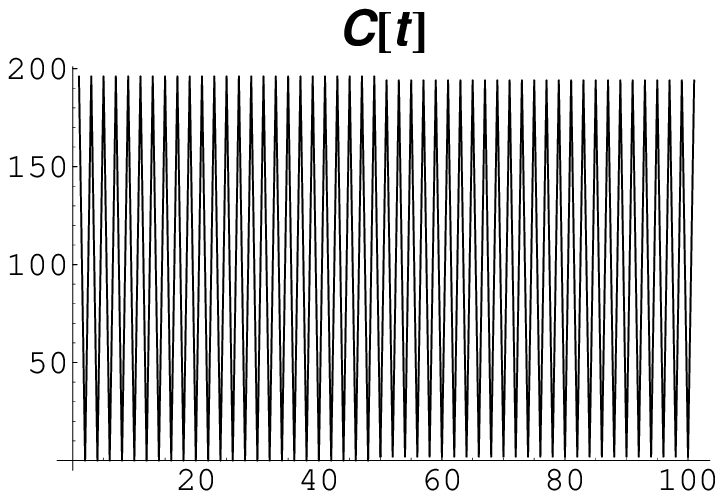}
\caption{Typical $P2$-state in $C(t)$ with $n=14$, $q=0.0005$ and $(P,T)=(-9,8)$.}
\end{center}
\end{figure}
In the peculiar points pointed out previously, the states $D1$ only appear.　These may give a deep understanding of a bifurcation phenomenon, as the latice size grows larger to be infinite degree of freedom. 　 

\subsection{p duality}
The reason why there is not any phase-transition like phenomenon exists mainly in the fact that the signs of Eqs. (2) and (6) are determined in the Chicken game.
This shows that an agent surrounded by all cooperaters or all defectors does not change the action by no means in any cases for various initial configurations of the Chicken game. 
So a checked pattern   consist of C and D is stable.  
Though the trendency to C or D depends on the parameter $P$ and $T$, the state 
necessarily trends to the checked pattern at $q \rightarrow \infty$. 
This is the reason why the state becomes $P1$ in $q \rightarrow \infty$ independent of various parameters and the initial states. 
They only reflects on the speed of the transition from $P1$ to $P2$. 
In all cases, however, that the stste transits through geometrical pattern to $P2$ can be understood analytically. 

As $q$ becomes smaller, $P1$ is disturbed and then at $q=0$ the all C and the all D states  repeat by turns in the cases of initial states (a)-(d).  
In the intermediate $q$, the trend toward converging on a  checked pattern competes with the one toward escaping from it stochastically and chaotic behaviors emerge. 

As we can also observed from  simulation results,  
there is the following symmetry with respect to $p$ which is the initial population ratio of C;   
\begin{eqnarray}
p &\Longleftrightarrow& 1-p \nonumber,\\
C &\Longleftrightarrow& D. 
\end{eqnarray}
We call this symmetry of the considering system $p\;\; duality$. 
This show that $p=0.5$ is a singular point, that is, a fixed point. 

From above considerations, the properties in the subsection 3.1 
result from  many facors involuved in complicate way.  
Next we investigate the effect of the factors in the subsequent subsections. 
First of all, the reflection of C and D arise only when C or D stands alone and is surrounded completely by the opposite action. 
What happens in the configuration that C or D is adjacency to each other?  
An extereme point where this happens is the fixed point $p=0.5$. 
This will be discussed in the next subsection 3.3. 

The order of a target agent is also essential for the above results. 
The case that the order of a target agent is taken on the cells at random is explored in 3.4.  
The situation that the size $n$ is even is needed for the stability of a checked pattern on Torus. When $n$ is taken as an odd number, a frustration prevents from  making a complete checked pattern. How about this case? 
We are going to study it in 3.5.
\subsection{Random initial configuration}
We discuss here the initial condition (e). 
the essential difference from (a) $\sim$ (d) arise when $D$ change into $P2$, where a new pattern $A$ appears (Fig. 11). 
So the transition pattern when $q$ becomes smaller is generally 
$$P1 \rightarrow C \rightarrow CI \rightarrow DI \rightarrow A \rightarrow P2,$$while for two singular points on the borderline it is 
  $$P1 \rightarrow C \rightarrow CI \rightarrow CID \rightarrow D \rightarrow A \rightarrow AP2 \rightarrow P2.$$
Though the state $A$ is similar to the state $D$ from appearances alone, 
it begins with the behavior like the  inverse of $D$ and $A-$like behavior is 
restored after a while (Fig. 11).  
When $q$ grows larger, the revival inverse $D$ like behavior spreads over the time series and it becomes to $P2$ in the end. 
$AP2$ is a middle state between $A$ and $p2$ (Fig. 12). 
Except for this point, there is not great difference between the behaviors with the randam initial configuration   and  the others in subsection 3.2. 
It is considered that the effect of mutual interference of an isolated C or D is not so drastic in general. 
The power of absorption into the checked pattern is so strong that there are not various roots in the transition from $P1$ to $P2$.   

\begin{figure}[t]
\begin{center}
\includegraphics[scale=1.0]{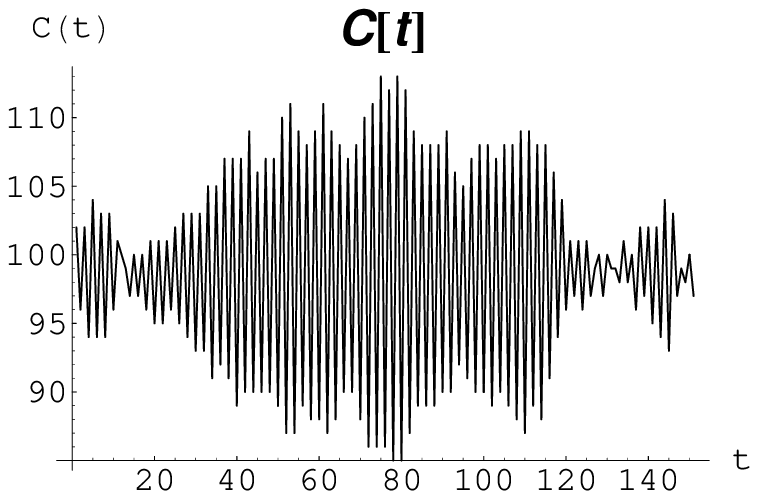}
\caption{Typical $A$-state in $C(t)$ with $n=14$, $q=0.001$ and $(P,T)=(-11,6)$.}
\end{center}
\end{figure}

\begin{figure}[t]
\begin{center}
\includegraphics[scale=1.0]{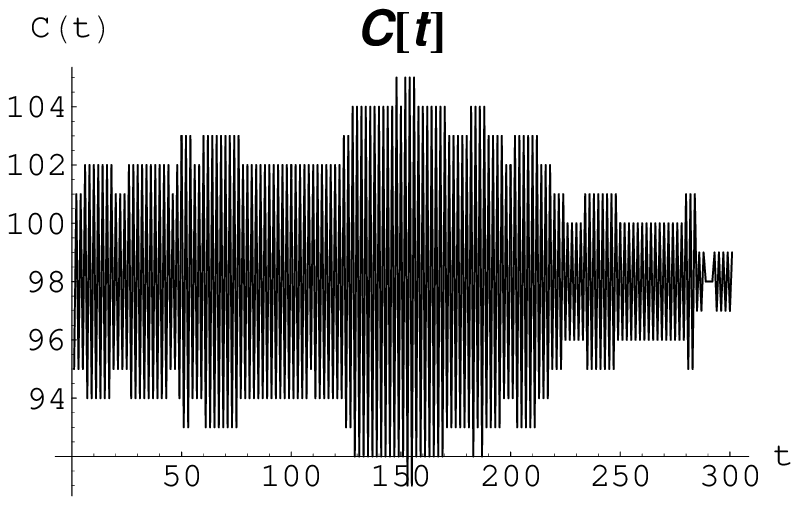}
\caption{Typical $AP2$-state in $C(t)$ with $n=14$, $q=0.0001$ and $(P,T)=(-11,6)$.}
\end{center}
\end{figure}

\subsection{Randum selection and P-T duality}
The results in the section 3.2 depends strongly on which agent is choosen first, that is, the order of a target agent. 
Here we study the case of a random ordering.  
Then the  dependence of $q$ in the transition is very simple. 
For all initial configuration, (a) $\sim$ (e), the transition is
$$P1 \rightarrow C.$$
In the Chicken game on the borderline, however, only $C$ appears. 

 In the simulation under (e), we find anther duality. 
At small $q$ in the Chicken game, the ratio of the population of C agents to total population in $P1$ is given by Table 2. This show that there is the following approximate duality; 
\begin{eqnarray}
Region \;\;I(II) &\Longleftrightarrow& Region\;\; IV(III), \nonumber \\
C &\Longleftrightarrow& D
\end{eqnarray} 

\begin{center}
Table 2. The ratio of the population of agents with the strategy C in each region.\\
\begin{tabular}{|c|c|c|c|c|}\hline
\makebox[15mm]{Region}&\makebox[10mm]{I}&\makebox[10mm]{II}&\makebox[10mm]{III}&\makebox[10mm]{IV}
\\ \hline 
\makebox[15mm]{Ratio of C}&\makebox[10mm]{0.65}&\makebox[10mm]{0.52}&\makebox[10mm]{0.45}&\makebox[10mm]{0.33}
\\ \hline 
\end{tabular}
\\
\end{center}

Since the regions in the Chicken game are specified by $P$ and $T$, we call this duality $P-T$ duality. This holds independently of the initial configurations and even on the borderline. 
This duality can be approximately understood by a simple analytic way but it is not  so nice to describe it due to the asynchronous time evolution.  
The analyses by the mean field method or a repricator equation \cite{Szab} may give better understanding for the duality but they are beyond this paper.

\subsection{Odd latice}
We discuss odd lattice, on which the checked pattern is not well defined and so a frustration arises. 
The simulation shows that the complete cheked pattern can not merely exist but the statistical properties  are invariant. 
As the size of the lattice grows larger, eventually these differences will disappear.  
Mainly we explore the case with initial condition (a). 
As a result, the states converge into $P1$ at $q\rightarrow \infty$ and the $P-T$ duality can be also observed in the population of C manifestly. 
A complete periodic pattern, however, curiously appears on the borderline between two areas. 
 An example is given in Fig. 13. 
Then we also observe the $P-T$ duality  in the periodic behavior itself.     

\begin{figure}[b]
\begin{center}
\includegraphics[scale=1.0]{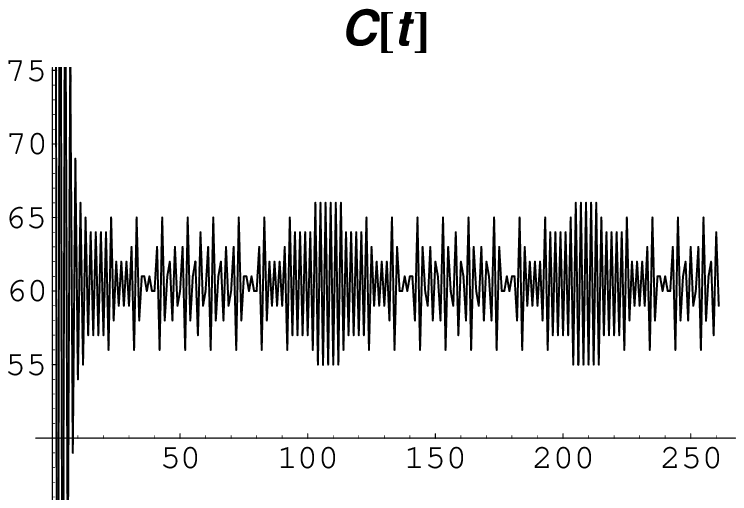}
\caption{ $C(t)$-state in the Chicken game  with $n=11$, and $(P,T)=(-8,9)$.}
\end{center}
\end{figure}

\section{Other Dilemma Games}
\subsection{Hero game and Leader game }
The Hero game and the Leader game also face a dilemma  because they have two Nash equilibria. 
All simulation results, however, are same as those of the Chicken game. 
Studying the details, the fact that these games also satisfy the Eqs. (2) and (6) seems to be crucial.  
We see that we neet only explore the Chicken game in a narrow sense but not study its relative games. 
Strangely  we observe that rough behaviors even in some non dilemmma games are same as those of the Chichen game.  

\subsection{Stag Hunt game}
This game face a dilemma in the sense that it has two Nash equilibria, but the one of which is a pareto optimal. 
It is not necessary that the relative games of the Stag Hunt game are explored in the similar  reason to the Chichen game. 

We simulate on 20 points based on the Fig. 4 for five kinds of initial state; (a)-(e).  
The points are taken from the regions I $\sim$ VI,  their border lines, I-II, II,III, III-IV, IV-V, V-VI, VI-I, and the border line between this game and its outside N, I-N, II-N, III-N, IV-N.   

We find a period two behavior for small $q$ where all C and all D appear alternately for (a) and (c) cases ( in the case (b) ((d)), one D (C) and all C (D) alternately repeat), and a convergence into one staste at large $q$ in $C(t)$. 
This properties is almost common to all initial configurations without the (e).  
In the intermediate $q$, some chaotic behaviors appeas like the Chicken game. 
The transition is the same as the $P1 \rightarrow C \rightarrow CI \rightarrow DI \rightarrow D \rightarrow P2.$

There, however, is one exception. It is the  peculiar point, at which three border lines intersect each other, in Fig. 4. 
The configuration oscillates between all C and all D all over $q$-values in this point.

There is one difference from the chicken in the convergent state, too. 
Though the convergent state in the Chicken game is a checked pattern, we find two convergent states in this game which depend on the regions showm in the Fig. 4.  
One is  a checked pattern same as in the Chicken game. 
Another is the all C (D) for the initial state (a) and (b) ((c) and (d)).
The former arises in areas II and III,  on their boundaries (I-II, II-III, III-IV) and on the border line between the this game and its outside N.    
The latter arises in all other regions.  
The signs of Eqs. (2) and (6) in this game are opposite to those of Chiken game.  
An agent tends to imitate the actions of the neighborhoods, that is, 
if the neighborhoods of the agent are all C(D), the agent takes C(D) strategy in the next step.   
Then a checked pattern, in which every C and D flip every inning, 
seems to be attractive. 
This occurs completely in regions II and III, where an agent abide by the decision of the majority (see Eqs. (3) and (5)). 
It is actually this time that the checked pattern arises.

It is considered that the  difference between the Chiken game and the Stag Hunt game is  due to the asynchronous  time evolution. 
Simulating under random ordering of a target agent in time evolution, the difference disappear. The both game show the same behaviors as the section 3.3.

In the case that begin with initial state (e),  three paterns appears at large $q$. 
The first pattern is that $C(t)$  fluctuates around the $N/2$ with period two. 
This occurs in the interval [I,III] where the symbol [,] means that the interval  includes  both ends. 
The second is to converge into all D state which  occurs in  the interval (IV,V) where the symbol (,) means that the interval does not  include  both ends. 
The third one is  to converge into all C state which occurs in the region IV, the border between V and IV  and the border between IV and I. 
On the border line between this game and its outside,  the first pattern appears basically. 
The behavior of $C(t)$ in the intermedidate $q$ is essentially the same as those in the Chicken game.  
At small $q$ ($q \rightarrow 0$), $C(t)$  fluctuates around the $N/2$.

\subsection{Prisoner's dilemma game}
In PD, as stated before, Eqs. (1)-(5) are trivially positive and there is not any sigunificant phase structure.  
The simulation results, however, are same as those of the Chiken game wholly.  
Only difference from the Chiken game is that $P1$ is all D because the action D is  a dominant action in PD. 
       
In the folowing subsection we investigate total hamiltonian in PD in  behalf of all game discussed in this paper. 

\subsection{Complexity and total hamiltonian}

We study the total hamiltonian in PD in which the essential properties do not depend on the explicit value of the parameters included in the payoff matrix. The total hamiltonian is  defined as the sum of the payoffs acquired by all agents at each step.  
This shows that how the behaviors of selfish agents have an influence on the whole system. 

The time series of the total hamiltonian and that of the population of C (or) is roughly alike (see Fig. 14). 
A little differense arise only in the chaotic phase. There is  an evident differences in Fig. 15. The population of C is increasing with (damped) oscillation during the initial (damped) oscillation in $C(t)$.  The reformation caused by the oscillation increase the total payoff in the system. 
\begin{figure}[h]
\begin{center}
\includegraphics[scale=0.7]{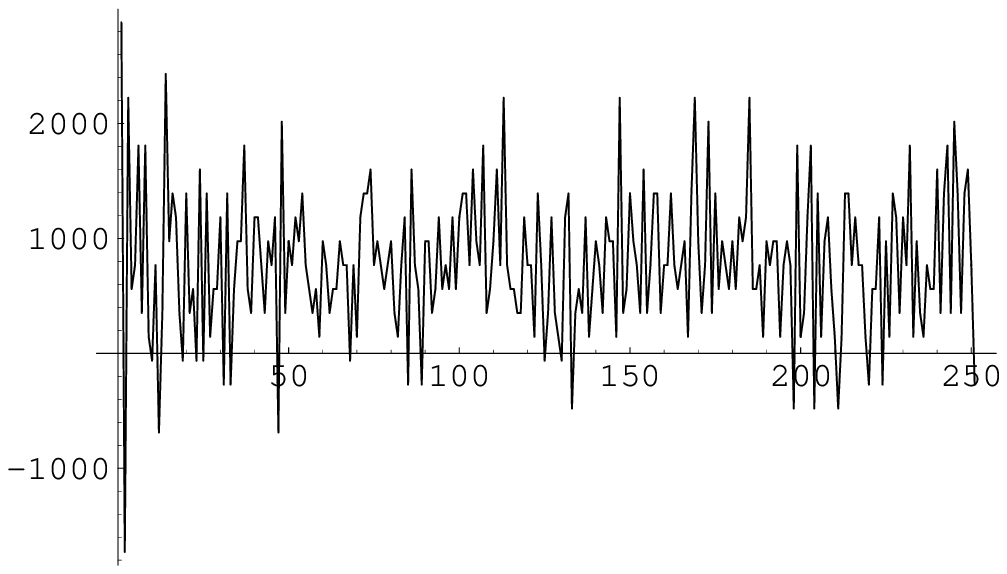}
\includegraphics[scale=0.7]{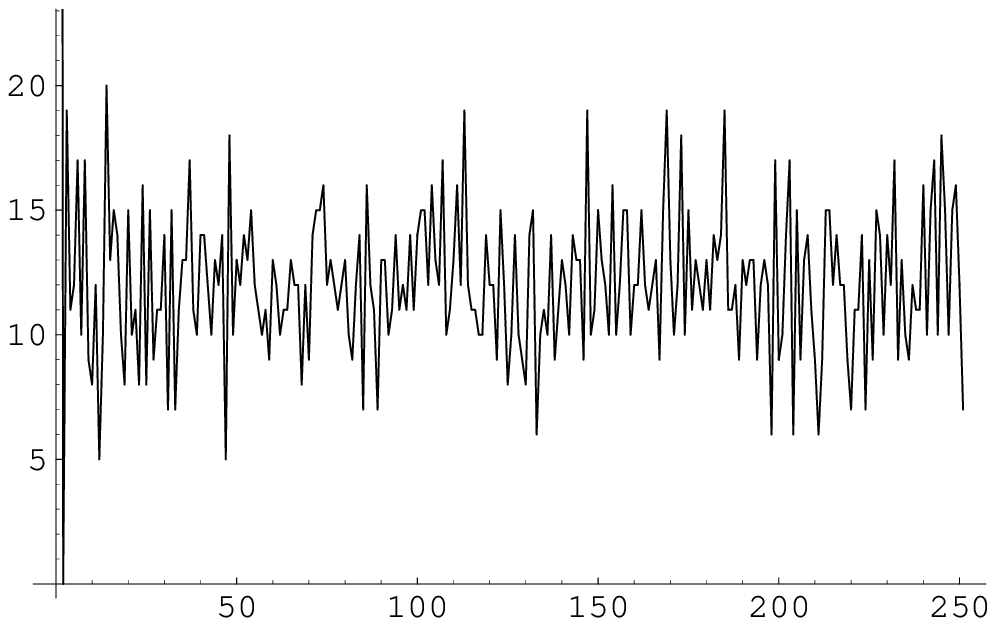}
\caption{ Total hamiltonian (left) and $C(t)$ (right) in the PD game with 
$(R,P,T,S)=(5,-3,50,-4)$ and $q=0.5)$.}
\end{center}
\end{figure}

\begin{figure}[h]
\begin{center}
\includegraphics[scale=0.7]{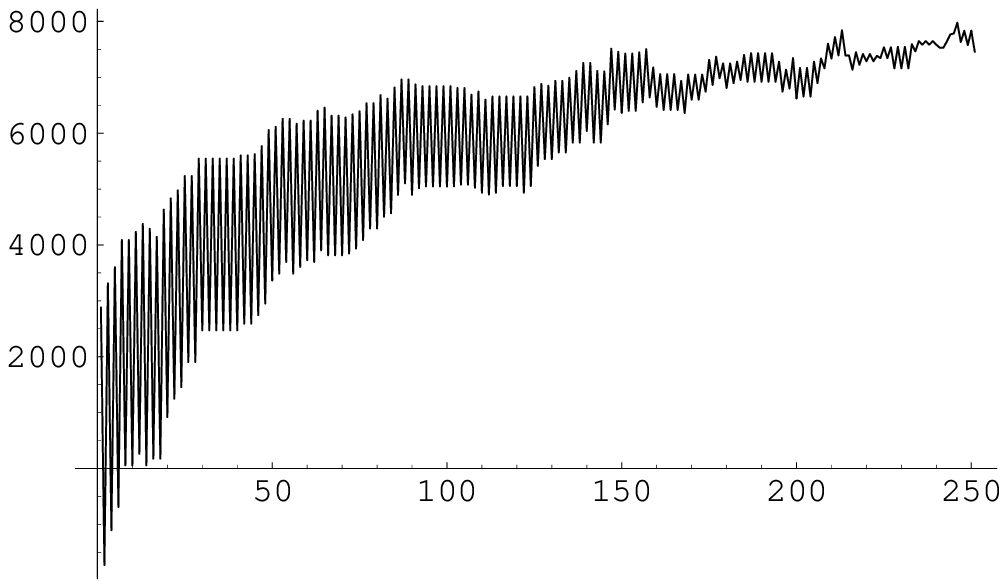}
\includegraphics[scale=0.7]{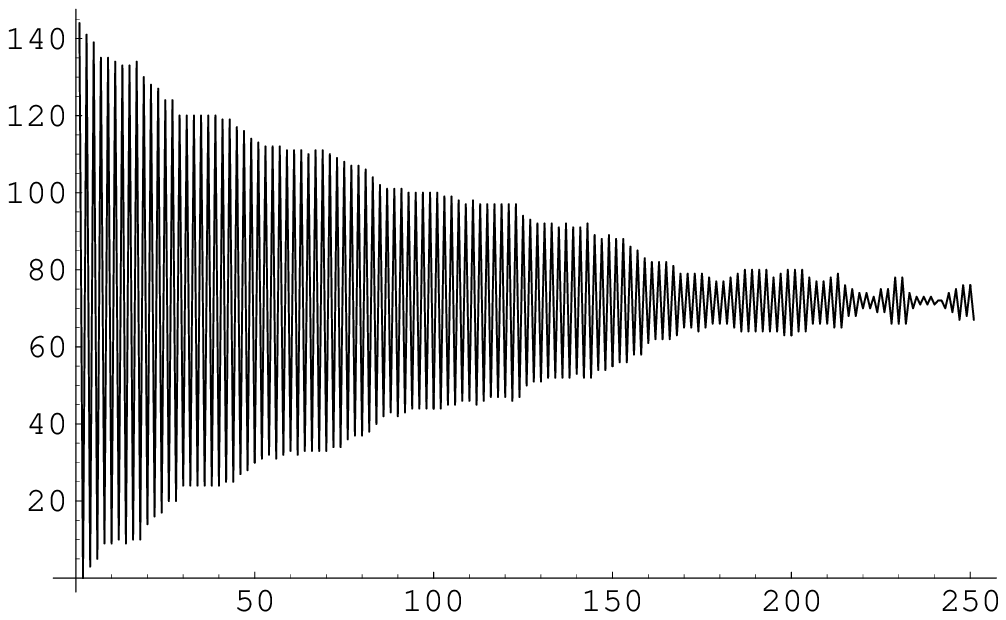}
\caption{ Total hamiltonian (above) and $C(t)$ (below) in the PD game with 
$(R,P,T,S)=(5,-3,50,-4)$ and $q=0.0001$.}
\end{center}
\end{figure}

 We estimate the time average of the total hamiltonian at each $q$ in PD.  
This is particularly significant when the configuration of the system is not convergent. 
Looking at Fig. 16, we find that the time average of the total hamiltonian 
reaches a maximum  in the chaotic phase. 
It is very suggestive that such complex disturbance induces the maximum of efficiency on an average. 
It may give the evolutionary understanding of the realtion between a complexity and the $\lambda$ parameter discussed by Langton \cite{Lang}.
The real maximum of the total hamiltonian is nearly equal to 7000, which is close to the convergent value in Fig. 15. 
Then the damping oscillation in $C(t)$ makes the total hamiltonian in the system  reach a potential maximum. 
As the oscillation becomes regularly periodic, that is to say small $q$,  the average value of the total hamiltonian decreases, again.  
In a complete two period where all C and all D arise repeatedly, the total hamiltonian   goes up and down between two values. The average is reduced to about 700 in PD. 
We should notice that  same results  apply  for the others games.  
\begin{figure}[h]
\begin{center}
\includegraphics[scale=1.0]{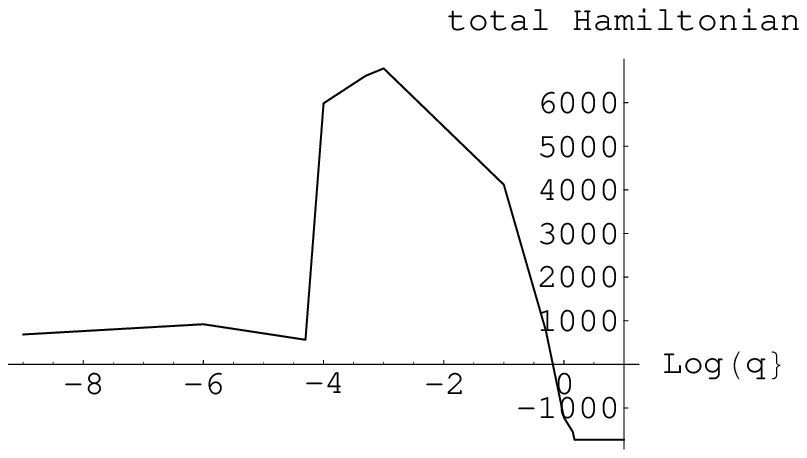}
\caption{ Total hamiltonian v.s. $\log q$ in the PD game.}
\end{center}
\end{figure}

\section{Concluding Remarks}
We study various types of spatial dilemma games under the evolution with the selfish rule on cells in this paper. 
Thus we find a kind of game universality, that is,  the  properties of time evolution  are game invariant in wide range.   
The behaviors show some common characteristics even in non-dilemma games. 

We must note that the explicit values in a payoff matrix themselves should play an important role, differently from the usual game theory. 
It leads to the varied phase diagrams (Figs.1-4), which satisfy the corresponding conditions made among $R$, $S$, $T$ and $P$ to the games. 
This is an important and theoretical defference between spatio-iterated games and non iterated games. 
The simulation results made under the selfish rule, however, mostly are the same as the Chicken game in a narrow sense. 
  
A common behavior of them are that a sort of  bifurcation is observed universally. 
Though we see one periodic begavior in $C(t)$ when a fluctuation parameter $q$ is large, as the parameters become smaller, $C(t)$ is chaotic and finally reaches a two period state at $q=0$. 
If a continuous limit of the infinite degrees of freedom is taken, the present analysis may lead to a deep understanding of a period double bifurcation or it also uncovers the difference between continous and descrete systems.  

We pointed out that in the chaotic phase the total hamiltonian reaches a  maximum. 
 This result should be compared with the one made under totalitarian strategy. 
The totalitarian may  not necessarily still more contribute to society  than egoist. Investigating this is so interesting and will be a next our work.



\end{document}

%% file: ChikinH2.tex
\unitlength 0.1in
\begin{picture}(63.30,25.40)(3.20,-28.40)
%
\special{pn 13}%
\special{pa 530 2300}%
\special{pa 3085 2310}%
\special{fp}%
\special{sh 1}%
\special{pa 3085 2310}%
\special{pa 3018 2290}%
\special{pa 3032 2310}%
\special{pa 3018 2330}%
\special{pa 3085 2310}%
\special{fp}%
%
\special{pn 13}%
\special{pa 2625 2410}%
\special{pa 2630 570}%
\special{fp}%
\special{sh 1}%
\special{pa 2630 570}%
\special{pa 2610 637}%
\special{pa 2630 623}%
\special{pa 2650 637}%
\special{pa 2630 570}%
\special{fp}%
%
\special{pn 8}%
\special{pa 630 1895}%
\special{pa 2935 1895}%
\special{fp}%
%
\special{pn 8}%
\special{pa 1825 515}%
\special{pa 2540 2460}%
\special{fp}%
%
\special{pn 8}%
\special{pa 904 630}%
\special{pa 2815 2310}%
\special{fp}%
%
\special{pn 8}%
\special{pa 620 1440}%
\special{pa 2800 2020}%
\special{fp}%
%
\special{pn 8}%
\special{pa 2335 505}%
\special{pa 2335 2365}%
\special{fp}%
\put(31.1000,-23.5000){\makebox(0,0)[lb]{P}}%
\put(26.0500,-5.6500){\makebox(0,0)[lb]{T}}%
\put(26.6500,-18.6500){\makebox(0,0)[lb]{5}}%
\put(22.1000,-25.5000){\makebox(0,0)[lb]{-4}}%
\put(18.1000,-20.8000){\makebox(0,0)[lb]{(-4,5)}}%
%
\special{pn 8}%
\special{pa 1070 1895}%
\special{pa 2340 995}%
\special{fp}%
\put(23.3000,-10.3500){\makebox(0,0)[lb]{(-4,13)}}%
\put(8.3000,-20.8000){\makebox(0,0)[lb]{(-12,5)}}%
\put(16.2000,-4.9000){\makebox(0,0)[lb]{(8)}}%
\put(6.0000,-4.9000){\makebox(0,0)[lb]{(9)}}%
\put(3.2000,-15.1000){\makebox(0,0)[lb]{(10)}}%
%
\special{pn 8}%
\special{sh 1}%
\special{ar 1415 1650 10 10 0  6.28318530717959E+0000}%
\special{sh 1}%
\special{ar 1415 1650 10 10 0  6.28318530717959E+0000}%
\special{sh 1}%
\special{ar 1415 1650 10 10 0  6.28318530717959E+0000}%
%
\special{pn 4}%
\special{sh 1}%
\special{ar 1075 1900 10 10 0  6.28318530717959E+0000}%
\special{sh 1}%
\special{ar 1060 1900 10 10 0  6.28318530717959E+0000}%
\special{sh 1}%
\special{ar 1060 1900 10 10 0  6.28318530717959E+0000}%
%
\special{pn 4}%
\special{sh 1}%
\special{ar 1410 1655 10 10 0  6.28318530717959E+0000}%
\special{sh 1}%
\special{ar 1410 1650 10 10 0  6.28318530717959E+0000}%
%
\special{pn 4}%
\special{sh 1}%
\special{ar 1775 1400 10 10 0  6.28318530717959E+0000}%
\special{sh 1}%
\special{ar 1780 1400 10 10 0  6.28318530717959E+0000}%
%
\special{pn 4}%
\special{sh 1}%
\special{ar 2075 1185 10 10 0  6.28318530717959E+0000}%
\special{sh 1}%
\special{ar 2070 1185 10 10 0  6.28318530717959E+0000}%
%
\special{pn 4}%
\special{sh 1}%
\special{ar 2330 1000 10 10 0  6.28318530717959E+0000}%
\special{sh 1}%
\special{ar 2330 1000 10 10 0  6.28318530717959E+0000}%
\special{sh 1}%
\special{ar 2330 1000 10 10 0  6.28318530717959E+0000}%
\put(15.1000,-12.4000){\makebox(0,0)[lb]{(-6,11)}}%
\put(12.5000,-14.8000){\makebox(0,0)[lb]{(-8,9)}}%
\put(8.2000,-17.5000){\makebox(0,0)[lb]{(-10,7)}}%
%
\special{pn 8}%
\special{pa 615 515}%
\special{pa 2340 510}%
\special{ip}%
%
\special{pn 8}%
\special{pa 630 515}%
\special{pa 645 1890}%
\special{ip}%
%
\special{pn 8}%
\special{pa 905 630}%
\special{pa 715 480}%
\special{fp}%
\put(20.5000,-7.8000){\makebox(0,0)[lb]{I}}%
\put(13.3000,-7.5000){\makebox(0,0)[lb]{II}}%
\put(6.8500,-10.6900){\makebox(0,0)[lb]{III}}%
\put(7.0000,-17.9000){\makebox(0,0)[lb]{IV}}%
%
\special{pn 4}%
\special{pa 1350 1020}%
\special{pa 870 1500}%
\special{fp}%
\special{pa 1410 1080}%
\special{pa 960 1530}%
\special{fp}%
\special{pa 1470 1140}%
\special{pa 1060 1550}%
\special{fp}%
\special{pa 1540 1190}%
\special{pa 1150 1580}%
\special{fp}%
\special{pa 1600 1250}%
\special{pa 1250 1600}%
\special{fp}%
\special{pa 1670 1300}%
\special{pa 1340 1630}%
\special{fp}%
\special{pa 1730 1360}%
\special{pa 1490 1600}%
\special{fp}%
\special{pa 1280 970}%
\special{pa 770 1480}%
\special{fp}%
\special{pa 1220 910}%
\special{pa 680 1450}%
\special{fp}%
\special{pa 1150 860}%
\special{pa 640 1370}%
\special{fp}%
\special{pa 1090 800}%
\special{pa 640 1250}%
\special{fp}%
\special{pa 1030 740}%
\special{pa 640 1130}%
\special{fp}%
\special{pa 960 690}%
\special{pa 640 1010}%
\special{fp}%
\special{pa 900 630}%
\special{pa 640 890}%
\special{fp}%
\special{pa 840 570}%
\special{pa 630 780}%
\special{fp}%
\special{pa 770 520}%
\special{pa 630 660}%
\special{fp}%
%
\special{pn 4}%
\special{pa 1870 620}%
\special{pa 1410 1080}%
\special{fp}%
\special{pa 1900 710}%
\special{pa 1480 1130}%
\special{fp}%
\special{pa 1930 800}%
\special{pa 1540 1190}%
\special{fp}%
\special{pa 1960 890}%
\special{pa 1610 1240}%
\special{fp}%
\special{pa 1990 980}%
\special{pa 1670 1300}%
\special{fp}%
\special{pa 2030 1060}%
\special{pa 1730 1360}%
\special{fp}%
\special{pa 2060 1150}%
\special{pa 1910 1300}%
\special{fp}%
\special{pa 1830 540}%
\special{pa 1350 1020}%
\special{fp}%
\special{pa 1740 510}%
\special{pa 1290 960}%
\special{fp}%
\special{pa 1620 510}%
\special{pa 1220 910}%
\special{fp}%
\special{pa 1500 510}%
\special{pa 1160 850}%
\special{fp}%
\special{pa 1370 520}%
\special{pa 1090 800}%
\special{fp}%
\special{pa 1250 520}%
\special{pa 1030 740}%
\special{fp}%
\special{pa 1130 520}%
\special{pa 970 680}%
\special{fp}%
\special{pa 1010 520}%
\special{pa 910 620}%
\special{fp}%
\special{pa 890 520}%
\special{pa 840 570}%
\special{fp}%
%
\special{pn 8}%
\special{pa 2340 750}%
\special{pa 2030 1060}%
\special{fp}%
\special{pa 2340 870}%
\special{pa 2060 1150}%
\special{fp}%
\special{pa 2340 630}%
\special{pa 2000 970}%
\special{fp}%
\special{pa 2330 520}%
\special{pa 1970 880}%
\special{fp}%
\special{pa 2220 510}%
\special{pa 1930 800}%
\special{fp}%
\special{pa 2100 510}%
\special{pa 1900 710}%
\special{fp}%
\special{pa 1980 510}%
\special{pa 1870 620}%
\special{fp}%
%
\special{pn 4}%
\special{pa 1050 1560}%
\special{pa 710 1900}%
\special{fp}%
\special{pa 1150 1580}%
\special{pa 830 1900}%
\special{fp}%
\special{pa 1240 1610}%
\special{pa 950 1900}%
\special{fp}%
\special{pa 1340 1630}%
\special{pa 1080 1890}%
\special{fp}%
\special{pa 960 1530}%
\special{pa 650 1840}%
\special{fp}%
\special{pa 860 1510}%
\special{pa 650 1720}%
\special{fp}%
\special{pa 770 1480}%
\special{pa 650 1600}%
\special{fp}%
\special{pa 670 1460}%
\special{pa 640 1490}%
\special{fp}%
%
\special{pn 4}%
\special{pa 1920 1530}%
\special{pa 1720 1730}%
\special{fp}%
\special{pa 1990 1580}%
\special{pa 1810 1760}%
\special{fp}%
\special{pa 2050 1640}%
\special{pa 1910 1780}%
\special{fp}%
\special{pa 2110 1700}%
\special{pa 2000 1810}%
\special{fp}%
\special{pa 2180 1750}%
\special{pa 2100 1830}%
\special{fp}%
\special{pa 2240 1810}%
\special{pa 2190 1860}%
\special{fp}%
\special{pa 1860 1470}%
\special{pa 1620 1710}%
\special{fp}%
\special{pa 1790 1420}%
\special{pa 1530 1680}%
\special{fp}%
%
\special{pn 4}%
\special{pa 2190 1500}%
\special{pa 2050 1640}%
\special{fp}%
\special{pa 2220 1590}%
\special{pa 2120 1690}%
\special{fp}%
\special{pa 2250 1680}%
\special{pa 2180 1750}%
\special{fp}%
\special{pa 2280 1770}%
\special{pa 2250 1800}%
\special{fp}%
\special{pa 2160 1410}%
\special{pa 1990 1580}%
\special{fp}%
\special{pa 2120 1330}%
\special{pa 1930 1520}%
\special{fp}%
\special{pa 2090 1240}%
\special{pa 1860 1470}%
\special{fp}%
\special{pa 1880 1330}%
\special{pa 1800 1410}%
\special{fp}%
%
\special{pn 4}%
\special{pa 2340 1350}%
\special{pa 2190 1500}%
\special{fp}%
\special{pa 2340 1470}%
\special{pa 2220 1590}%
\special{fp}%
\special{pa 2340 1590}%
\special{pa 2250 1680}%
\special{fp}%
\special{pa 2340 1710}%
\special{pa 2290 1760}%
\special{fp}%
\special{pa 2340 1830}%
\special{pa 2320 1850}%
\special{fp}%
\special{pa 2340 1230}%
\special{pa 2160 1410}%
\special{fp}%
\special{pa 2340 1110}%
\special{pa 2130 1320}%
\special{fp}%
\special{pa 2290 1040}%
\special{pa 2090 1240}%
\special{fp}%
%
\special{pn 4}%
\special{pa 1620 1710}%
\special{pa 1430 1900}%
\special{fp}%
\special{pa 1530 1680}%
\special{pa 1310 1900}%
\special{fp}%
\special{pa 1430 1660}%
\special{pa 1190 1900}%
\special{fp}%
\special{pa 1720 1730}%
\special{pa 1550 1900}%
\special{fp}%
\special{pa 1810 1760}%
\special{pa 1670 1900}%
\special{fp}%
\special{pa 1910 1780}%
\special{pa 1790 1900}%
\special{fp}%
\special{pa 2000 1810}%
\special{pa 1910 1900}%
\special{fp}%
\special{pa 2100 1830}%
\special{pa 2030 1900}%
\special{fp}%
\special{pa 2190 1860}%
\special{pa 2150 1900}%
\special{fp}%
%
\special{pn 13}%
\special{pa 3640 870}%
\special{pa 6650 890}%
\special{fp}%
\special{sh 1}%
\special{pa 6650 890}%
\special{pa 6583 870}%
\special{pa 6597 890}%
\special{pa 6583 910}%
\special{pa 6650 890}%
\special{fp}%
%
\special{pn 13}%
\special{pa 3705 2275}%
\special{pa 3705 415}%
\special{fp}%
\special{sh 1}%
\special{pa 3705 415}%
\special{pa 3685 482}%
\special{pa 3705 468}%
\special{pa 3725 482}%
\special{pa 3705 415}%
\special{fp}%
%
\special{pn 8}%
\special{pa 3560 670}%
\special{pa 6085 670}%
\special{da 0.070}%
%
\special{pn 8}%
\special{pa 5225 400}%
\special{pa 5225 2170}%
\special{da 0.070}%
\put(50.2000,-6.3000){\makebox(0,0)[lb]{(T,R)}}%
\put(37.2500,-6.5000){\makebox(0,0)[lb]{R}}%
\put(51.4000,-10.7000){\makebox(0,0)[lb]{T}}%
%
\special{pn 8}%
\special{pa 3550 1350}%
\special{pa 4545 485}%
\special{fp}%
%
\special{pn 8}%
\special{pa 5420 500}%
\special{pa 3575 2080}%
\special{fp}%
%
\special{pn 8}%
\special{pa 5990 505}%
\special{pa 3930 2360}%
\special{fp}%
\put(37.0000,-13.0000){\makebox(0,0)[lb]{-a/3}}%
\put(41.9500,-8.6000){\makebox(0,0)[lb]{a/3}}%
\put(37.2000,-20.3500){\makebox(0,0)[lb]{-a  }}%
\put(49.9000,-10.1000){\makebox(0,0)[lb]{a}}%
\put(54.9500,-8.7000){\makebox(0,0)[lb]{3a}}%
\put(50.0500,-12.4000){\makebox(0,0)[lb]{T-3a}}%
\put(54.4000,-6.5000){\makebox(0,0)[lb]{(R+3a,R)}}%
\put(39.7000,-6.6000){\makebox(0,0)[lb]{(R+a/3,R)}}%
%
\special{pn 4}%
\special{pa 6120 890}%
\special{pa 5225 2145}%
\special{fp}%
%
\special{pn 4}%
\special{pa 5235 2110}%
\special{pa 6095 2095}%
\special{ip}%
\special{pa 6090 2105}%
\special{pa 6095 2110}%
\special{ip}%
%
\special{pn 8}%
\special{pa 6100 895}%
\special{pa 6080 2095}%
\special{ip}%
%
\special{pn 8}%
\special{pa 6080 2010}%
\special{pa 5640 1570}%
\special{fp}%
\special{pa 6080 1890}%
\special{pa 5690 1500}%
\special{fp}%
\special{pa 6090 1780}%
\special{pa 5740 1430}%
\special{fp}%
\special{pa 6090 1660}%
\special{pa 5790 1360}%
\special{fp}%
\special{pa 6090 1540}%
\special{pa 5840 1290}%
\special{fp}%
\special{pa 6090 1420}%
\special{pa 5890 1220}%
\special{fp}%
\special{pa 6090 1300}%
\special{pa 5940 1150}%
\special{fp}%
\special{pa 6100 1190}%
\special{pa 5990 1080}%
\special{fp}%
\special{pa 6100 1070}%
\special{pa 6040 1010}%
\special{fp}%
\special{pa 6050 2100}%
\special{pa 5590 1640}%
\special{fp}%
\special{pa 5930 2100}%
\special{pa 5540 1710}%
\special{fp}%
\special{pa 5810 2100}%
\special{pa 5490 1780}%
\special{fp}%
\special{pa 5690 2100}%
\special{pa 5440 1850}%
\special{fp}%
\special{pa 5580 2110}%
\special{pa 5390 1920}%
\special{fp}%
\special{pa 5460 2110}%
\special{pa 5340 1990}%
\special{fp}%
\special{pa 5340 2110}%
\special{pa 5290 2060}%
\special{fp}%
%
\special{pn 4}%
\special{pa 5590 1640}%
\special{pa 5230 1280}%
\special{fp}%
\special{pa 5640 1570}%
\special{pa 5250 1180}%
\special{fp}%
\special{pa 5690 1500}%
\special{pa 5310 1120}%
\special{fp}%
\special{pa 5740 1430}%
\special{pa 5380 1070}%
\special{fp}%
\special{pa 5790 1360}%
\special{pa 5440 1010}%
\special{fp}%
\special{pa 5840 1290}%
\special{pa 5500 950}%
\special{fp}%
\special{pa 5890 1220}%
\special{pa 5570 900}%
\special{fp}%
\special{pa 5940 1150}%
\special{pa 5670 880}%
\special{fp}%
\special{pa 5990 1080}%
\special{pa 5790 880}%
\special{fp}%
\special{pa 6040 1010}%
\special{pa 5910 880}%
\special{fp}%
\special{pa 5540 1710}%
\special{pa 5230 1400}%
\special{fp}%
\special{pa 5490 1780}%
\special{pa 5230 1520}%
\special{fp}%
\special{pa 5440 1850}%
\special{pa 5230 1640}%
\special{fp}%
\special{pa 5390 1920}%
\special{pa 5230 1760}%
\special{fp}%
\special{pa 5340 1990}%
\special{pa 5230 1880}%
\special{fp}%
\special{pa 5290 2060}%
\special{pa 5230 2000}%
\special{fp}%
%
\special{pn 4}%
\special{pa 5370 1060}%
\special{pa 5230 920}%
\special{fp}%
\special{pa 5430 1000}%
\special{pa 5310 880}%
\special{fp}%
\special{pa 5500 950}%
\special{pa 5430 880}%
\special{fp}%
\special{pa 5310 1120}%
\special{pa 5230 1040}%
\special{fp}%
\put(5.5000,-27.9000){\makebox(0,0)[lb]{Fig1. Phase diagram of the Chikin game}}%
\put(5.3000,-30.1000){\makebox(0,0)[lb]{with the selfish evolutionary rule.}}%
\put(39.0000,-28.1000){\makebox(0,0)[lb]{Fig2. Phase diagram of the Hero game}}%
\put(40.0000,-30.1000){\makebox(0,0)[lb]{with the selfish evolutionary rule.}}%
\put(33.2000,-15.2000){\makebox(0,0)[lb]{(13)}}%
\put(33.5000,-20.6000){\makebox(0,0)[lb]{(12)}}%
\put(36.3000,-24.4000){\makebox(0,0)[lb]{(11)}}%
\put(66.0000,-11.1000){\makebox(0,0)[lb]{S}}%
\put(36.3000,-4.7000){\makebox(0,0)[lb]{P}}%
\end{picture}%

%% file: LeaderSt6-3.tex
\unitlength 0.1in
\begin{picture}(65.30,33.60)(0.00,-29.60)
%
\special{pn 8}%
\special{pa 1360 490}%
\special{pa 2295 1705}%
\special{fp}%
%
\special{pn 13}%
\special{pa 445 2355}%
\special{pa 2510 2345}%
\special{fp}%
\special{sh 1}%
\special{pa 2510 2345}%
\special{pa 2443 2325}%
\special{pa 2457 2345}%
\special{pa 2443 2365}%
\special{pa 2510 2345}%
\special{fp}%
%
\special{pn 13}%
\special{pa 2230 2480}%
\special{pa 2230 625}%
\special{fp}%
\special{sh 1}%
\special{pa 2230 625}%
\special{pa 2210 692}%
\special{pa 2230 678}%
\special{pa 2250 692}%
\special{pa 2230 625}%
\special{fp}%
%
\special{pn 8}%
\special{pa 475 1830}%
\special{pa 2520 1830}%
\special{fp}%
%
\special{pn 8}%
\special{pa 1780 620}%
\special{pa 1780 2525}%
\special{fp}%
%
\special{pn 8}%
\special{pa 515 1730}%
\special{pa 1710 1730}%
\special{da 0.070}%
%
\special{pn 8}%
\special{pa 1685 1730}%
\special{pa 1685 760}%
\special{da 0.070}%
%
\special{pn 8}%
\special{pa 790 940}%
\special{pa 2430 2395}%
\special{fp}%
\put(22.6000,-18.3000){\makebox(0,0)[lb]{5}}%
\put(17.1000,-24.4000){\makebox(0,0)[lb]{-4}}%
\put(14.2000,-18.2000){\makebox(0,0)[lb]{(-5,6)}}%
\put(13.7000,-9.9000){\makebox(0,0)[lb]{(-5,26)}}%
\put(8.7000,-17.1500){\makebox(0,0)[lb]{(-25,6)}}%
%
\special{pn 4}%
\special{sh 1}%
\special{ar 910 1725 10 10 0  6.28318530717959E+0000}%
\special{sh 1}%
\special{ar 915 1730 10 10 0  6.28318530717959E+0000}%
\special{sh 1}%
\special{ar 915 1730 10 10 0  6.28318530717959E+0000}%
\special{sh 1}%
\special{ar 910 1725 10 10 0  6.28318530717959E+0000}%
\special{sh 1}%
\special{ar 910 1725 10 10 0  6.28318530717959E+0000}%
%
\special{pn 4}%
\special{sh 1}%
\special{ar 1685 1735 10 10 0  6.28318530717959E+0000}%
\special{sh 1}%
\special{ar 1680 1730 10 10 0  6.28318530717959E+0000}%
\special{sh 1}%
\special{ar 1680 1730 10 10 0  6.28318530717959E+0000}%
%
\special{pn 4}%
\special{sh 1}%
\special{ar 1670 670 10 10 0  6.28318530717959E+0000}%
\special{sh 1}%
\special{ar 1680 915 10 10 0  6.28318530717959E+0000}%
\special{sh 1}%
\special{ar 1680 915 10 10 0  6.28318530717959E+0000}%
%
\special{pn 8}%
\special{pa 470 600}%
\special{pa 485 1855}%
\special{fp}%
%
\special{pn 8}%
\special{pa 390 690}%
\special{pa 1860 700}%
\special{fp}%
%
\special{pn 8}%
\special{pa 785 935}%
\special{pa 494 666}%
\special{fp}%
\put(21.9000,-5.7000){\makebox(0,0)[lb]{T}}%
\put(26.1000,-24.8000){\makebox(0,0)[lb]{P}}%
%
\special{pn 4}%
\special{pa 1470 1730}%
\special{pa 1370 1830}%
\special{fp}%
\special{pa 1350 1730}%
\special{pa 1250 1830}%
\special{fp}%
\special{pa 1230 1730}%
\special{pa 1130 1830}%
\special{fp}%
\special{pa 1110 1730}%
\special{pa 1010 1830}%
\special{fp}%
\special{pa 990 1730}%
\special{pa 890 1830}%
\special{fp}%
\special{pa 870 1730}%
\special{pa 770 1830}%
\special{fp}%
\special{pa 750 1730}%
\special{pa 650 1830}%
\special{fp}%
\special{pa 630 1730}%
\special{pa 530 1830}%
\special{fp}%
\special{pa 1060 1180}%
\special{pa 490 1750}%
\special{fp}%
\special{pa 990 1130}%
\special{pa 490 1630}%
\special{fp}%
\special{pa 930 1070}%
\special{pa 480 1520}%
\special{fp}%
\special{pa 870 1010}%
\special{pa 480 1400}%
\special{fp}%
\special{pa 800 960}%
\special{pa 480 1280}%
\special{fp}%
\special{pa 740 900}%
\special{pa 480 1160}%
\special{fp}%
\special{pa 680 840}%
\special{pa 480 1040}%
\special{fp}%
\special{pa 610 790}%
\special{pa 480 920}%
\special{fp}%
\special{pa 550 730}%
\special{pa 470 810}%
\special{fp}%
\special{pa 1120 1240}%
\special{pa 630 1730}%
\special{fp}%
\special{pa 1180 1300}%
\special{pa 750 1730}%
\special{fp}%
\special{pa 1250 1350}%
\special{pa 870 1730}%
\special{fp}%
\special{pa 1310 1410}%
\special{pa 990 1730}%
\special{fp}%
\special{pa 1370 1470}%
\special{pa 1110 1730}%
\special{fp}%
\special{pa 1440 1520}%
\special{pa 1230 1730}%
\special{fp}%
\special{pa 1500 1580}%
\special{pa 1350 1730}%
\special{fp}%
\special{pa 1570 1630}%
\special{pa 1470 1730}%
\special{fp}%
\special{pa 1630 1690}%
\special{pa 1590 1730}%
\special{fp}%
\special{pa 1590 1730}%
\special{pa 1490 1830}%
\special{fp}%
\special{pa 1690 1750}%
\special{pa 1610 1830}%
\special{fp}%
%
\special{pn 4}%
\special{pa 1780 1300}%
\special{pa 1690 1390}%
\special{fp}%
\special{pa 1780 1420}%
\special{pa 1690 1510}%
\special{fp}%
\special{pa 1780 1540}%
\special{pa 1690 1630}%
\special{fp}%
\special{pa 1780 1660}%
\special{pa 1700 1740}%
\special{fp}%
\special{pa 1780 1180}%
\special{pa 1690 1270}%
\special{fp}%
\special{pa 1780 1060}%
\special{pa 1690 1150}%
\special{fp}%
\special{pa 1740 980}%
\special{pa 1690 1030}%
\special{fp}%
%
\special{pn 4}%
\special{pa 1420 700}%
\special{pa 1000 1120}%
\special{fp}%
\special{pa 1530 710}%
\special{pa 1060 1180}%
\special{fp}%
\special{pa 1580 780}%
\special{pa 1130 1230}%
\special{fp}%
\special{pa 1630 850}%
\special{pa 1190 1290}%
\special{fp}%
\special{pa 1670 930}%
\special{pa 1250 1350}%
\special{fp}%
\special{pa 1690 1030}%
\special{pa 1320 1400}%
\special{fp}%
\special{pa 1690 1150}%
\special{pa 1380 1460}%
\special{fp}%
\special{pa 1690 1270}%
\special{pa 1440 1520}%
\special{fp}%
\special{pa 1690 1390}%
\special{pa 1510 1570}%
\special{fp}%
\special{pa 1690 1510}%
\special{pa 1570 1630}%
\special{fp}%
\special{pa 1690 1630}%
\special{pa 1630 1690}%
\special{fp}%
\special{pa 1300 700}%
\special{pa 940 1060}%
\special{fp}%
\special{pa 1180 700}%
\special{pa 870 1010}%
\special{fp}%
\special{pa 1070 690}%
\special{pa 810 950}%
\special{fp}%
\special{pa 950 690}%
\special{pa 740 900}%
\special{fp}%
\special{pa 830 690}%
\special{pa 680 840}%
\special{fp}%
\special{pa 710 690}%
\special{pa 620 780}%
\special{fp}%
\put(1.0000,-28.8000){\makebox(0,0)[lb]{Fig3. Phase structure of the Leader game}}%
\put(3.4000,-31.3000){\makebox(0,0)[lb]{with selfish evolutionary rule.}}%
%
\special{pn 20}%
\special{pa 3520 2220}%
\special{pa 5950 2229}%
\special{fp}%
\special{sh 1}%
\special{pa 5950 2229}%
\special{pa 5883 2209}%
\special{pa 5897 2229}%
\special{pa 5883 2249}%
\special{pa 5950 2229}%
\special{fp}%
%
\special{pn 20}%
\special{pa 5644 2598}%
\special{pa 5644 708}%
\special{fp}%
\special{sh 1}%
\special{pa 5644 708}%
\special{pa 5624 775}%
\special{pa 5644 761}%
\special{pa 5664 775}%
\special{pa 5644 708}%
\special{fp}%
%
\special{pn 4}%
\special{ar 4132 1338 40 40  0.9272952 0.9827937}%
%
\special{pn 4}%
\special{ar 4123 1374 38 38  1.8637087 1.8660320}%
%
\special{pn 4}%
\special{sh 0.600}%
\special{ar 4285 1356 30 35  0.0000000 6.2831853}%
%
\special{pn 4}%
\special{pa 6229 2220}%
\special{pa 3368 943}%
\special{fp}%
%
\special{pn 8}%
\special{pa 3313 1114}%
\special{pa 6402 1916}%
\special{fp}%
%
\special{pn 8}%
\special{pa 3637 771}%
\special{pa 5698 2634}%
\special{fp}%
\special{pa 4159 1257}%
\special{pa 4150 1230}%
\special{fp}%
\special{pa 4141 1185}%
\special{pa 4123 1158}%
\special{fp}%
\special{pa 4096 1185}%
\special{pa 4096 1194}%
\special{fp}%
\special{pa 4096 1194}%
\special{pa 4159 1239}%
\special{fp}%
\special{pa 4159 1239}%
\special{pa 4159 1239}%
\special{fp}%
%
\special{pn 4}%
\special{pa 5185 2751}%
\special{pa 6319 1482}%
\special{fp}%
\put(56.2600,-7.0800){\makebox(0,0)[lb]{R}}%
\put(60.2200,-23.3700){\makebox(0,0)[lb]{S}}%
\put(49.6900,-12.4800){\makebox(0,0)[lb]{(I)}}%
\put(50.8600,-17.6100){\makebox(0,0)[lb]{(II)}}%
\put(48.3400,-19.0500){\makebox(0,0)[lb]{(III)}}%
\put(37.0900,-19.9500){\makebox(0,0)[lb]{(IV)}}%
\put(31.6900,-11.2200){\makebox(0,0)[lb]{(V)}}%
\put(33.4000,-9.1500){\makebox(0,0)[lb]{(VI)}}%
\put(58.2400,-17.5200){\makebox(0,0)[lb]{(3,3)}}%
\put(58.4200,-21.1200){\makebox(0,0)[lb]{(1,1)}}%
\put(49.1500,-25.2600){\makebox(0,0)[lb]{(-1,-1)}}%
\put(38.8900,-15.5400){\makebox(0,0)[lb]{(-3,5)}}%
%
\special{pn 8}%
\special{pa 3151 1680}%
\special{pa 4312 537}%
\special{dt 0.045}%
\special{pa 4312 537}%
\special{pa 4312 537}%
\special{dt 0.045}%
%
\special{pn 8}%
\special{pa 4096 2400}%
\special{pa 5743 1005}%
\special{dt 0.045}%
\special{pa 5743 1005}%
\special{pa 5742 1005}%
\special{dt 0.045}%
\put(51.4000,-16.0800){\makebox(0,0)[lb]{(0,4)}}%
\put(45.7300,-17.6100){\makebox(0,0)[lb]{(-1,3)}}%
\put(42.8500,-19.1400){\makebox(0,0)[lb]{(-2,2)}}%
\put(39.6100,-10.4100){\makebox(0,0)[lb]{(-4,8)}}%
\put(30.8800,-13.6500){\makebox(0,0)[lb]{(-6,6)}}%
\put(34.9300,-11.2200){\makebox(0,0)[lb]{(-5,7)}}%
%
\special{pn 8}%
\special{pa 0 7007188}%
\special{sp}%
%
\special{pn 8}%
\special{pa 0 7007188}%
\special{sp}%
%
\special{pn 8}%
\special{pa 5392 2769}%
\special{pa 2782 1950}%
\special{ip}%
%
\special{pn 8}%
\special{pa 6310 1662}%
\special{pa 5887 285}%
\special{ip}%
%
\special{pn 8}%
\special{pa 3493 492}%
\special{pa 3520 2328}%
\special{ip}%
%
\special{pn 4}%
\special{pa 5608 1122}%
\special{pa 4024 1122}%
\special{fp}%
\special{pa 5635 1041}%
\special{pa 3934 1041}%
\special{fp}%
\special{pa 5635 960}%
\special{pa 3880 960}%
\special{fp}%
\special{pa 5635 879}%
\special{pa 3961 879}%
\special{fp}%
\special{pa 5635 798}%
\special{pa 4051 798}%
\special{fp}%
\special{pa 5626 717}%
\special{pa 4132 717}%
\special{fp}%
\special{pa 5995 636}%
\special{pa 4213 636}%
\special{fp}%
\special{pa 6022 717}%
\special{pa 5662 717}%
\special{fp}%
\special{pa 6049 798}%
\special{pa 5653 798}%
\special{fp}%
\special{pa 6067 879}%
\special{pa 5653 879}%
\special{fp}%
\special{pa 6094 960}%
\special{pa 5653 960}%
\special{fp}%
\special{pa 5698 1041}%
\special{pa 5653 1041}%
\special{fp}%
\special{pa 6121 1041}%
\special{pa 5698 1041}%
\special{fp}%
\special{pa 6148 1122}%
\special{pa 5653 1122}%
\special{fp}%
\special{pa 6166 1203}%
\special{pa 5653 1203}%
\special{fp}%
\special{pa 6193 1284}%
\special{pa 5653 1284}%
\special{fp}%
\special{pa 6220 1365}%
\special{pa 5653 1365}%
\special{fp}%
\special{pa 6247 1446}%
\special{pa 5653 1446}%
\special{fp}%
\special{pa 6265 1527}%
\special{pa 5653 1527}%
\special{fp}%
\special{pa 6202 1608}%
\special{pa 5653 1608}%
\special{fp}%
\special{pa 6130 1689}%
\special{pa 5653 1689}%
\special{fp}%
\special{pa 6058 1770}%
\special{pa 5860 1770}%
\special{fp}%
\special{pa 5509 1203}%
\special{pa 4114 1203}%
\special{fp}%
\special{pa 5410 1284}%
\special{pa 4204 1284}%
\special{fp}%
\special{pa 5320 1365}%
\special{pa 4321 1365}%
\special{fp}%
\special{pa 5221 1446}%
\special{pa 4609 1446}%
\special{fp}%
\special{pa 5122 1527}%
\special{pa 4924 1527}%
\special{fp}%
%
\special{pn 4}%
\special{pa 4132 717}%
\special{pa 3493 717}%
\special{fp}%
\special{pa 4213 636}%
\special{pa 3493 636}%
\special{fp}%
\special{pa 3664 798}%
\special{pa 3502 798}%
\special{fp}%
\special{pa 3754 879}%
\special{pa 3502 879}%
\special{fp}%
\special{pa 3844 960}%
\special{pa 3502 960}%
\special{fp}%
\special{pa 3799 1041}%
\special{pa 3592 1041}%
\special{fp}%
\special{pa 4051 798}%
\special{pa 3664 798}%
\special{fp}%
\special{pa 3961 879}%
\special{pa 3754 879}%
\special{fp}%
\special{pa 3880 960}%
\special{pa 3844 960}%
\special{fp}%
%
\special{pn 4}%
\special{pa 4024 1122}%
\special{pa 3772 1122}%
\special{fp}%
\special{pa 3934 1041}%
\special{pa 3799 1041}%
\special{fp}%
\special{pa 4105 1203}%
\special{pa 3961 1203}%
\special{fp}%
\special{pa 4204 1284}%
\special{pa 4141 1284}%
\special{fp}%
%
\special{pn 4}%
\special{pa 3943 1203}%
\special{pa 3673 1203}%
\special{fp}%
\special{pa 3763 1122}%
\special{pa 3718 1122}%
\special{fp}%
%
\special{pn 4}%
\special{pa 3718 1122}%
\special{pa 3502 1122}%
\special{fp}%
\special{pa 3583 1041}%
\special{pa 3502 1041}%
\special{fp}%
%
\special{pn 4}%
\special{pa 4555 2013}%
\special{pa 3511 2013}%
\special{fp}%
\special{pa 4645 1932}%
\special{pa 3511 1932}%
\special{fp}%
\special{pa 4744 1851}%
\special{pa 3511 1851}%
\special{fp}%
\special{pa 4744 1770}%
\special{pa 3511 1770}%
\special{fp}%
\special{pa 4654 1689}%
\special{pa 3511 1689}%
\special{fp}%
\special{pa 4564 1608}%
\special{pa 3511 1608}%
\special{fp}%
\special{pa 4474 1527}%
\special{pa 3511 1527}%
\special{fp}%
\special{pa 4384 1446}%
\special{pa 3511 1446}%
\special{fp}%
\special{pa 4258 1365}%
\special{pa 3502 1365}%
\special{fp}%
\special{pa 3961 1284}%
\special{pa 3556 1284}%
\special{fp}%
\special{pa 4456 2094}%
\special{pa 3520 2094}%
\special{fp}%
\special{pa 4357 2175}%
\special{pa 3520 2175}%
\special{fp}%
%
\special{pn 4}%
\special{pa 5410 2499}%
\special{pa 4537 2499}%
\special{fp}%
\special{pa 5455 2418}%
\special{pa 4285 2418}%
\special{fp}%
\special{pa 5365 2337}%
\special{pa 4168 2337}%
\special{fp}%
\special{pa 5275 2256}%
\special{pa 4267 2256}%
\special{fp}%
\special{pa 5338 2580}%
\special{pa 4798 2580}%
\special{fp}%
\special{pa 5266 2661}%
\special{pa 5059 2661}%
\special{fp}%
%
\special{pn 4}%
\special{pa 5104 2094}%
\special{pa 4456 2094}%
\special{fp}%
\special{pa 5194 2175}%
\special{pa 4366 2175}%
\special{fp}%
\special{pa 5014 2013}%
\special{pa 4555 2013}%
\special{fp}%
\special{pa 4924 1932}%
\special{pa 4645 1932}%
\special{fp}%
\special{pa 4834 1851}%
\special{pa 4744 1851}%
\special{fp}%
%
\special{pn 4}%
\special{pa 5635 2013}%
\special{pa 5014 2013}%
\special{fp}%
\special{pa 5581 1932}%
\special{pa 4924 1932}%
\special{fp}%
\special{pa 5401 1851}%
\special{pa 4834 1851}%
\special{fp}%
\special{pa 5212 1770}%
\special{pa 4843 1770}%
\special{fp}%
\special{pa 5032 1689}%
\special{pa 4933 1689}%
\special{fp}%
\special{pa 5635 2094}%
\special{pa 5104 2094}%
\special{fp}%
\special{pa 5635 2175}%
\special{pa 5194 2175}%
\special{fp}%
%
\special{pn 4}%
\special{pa 5635 1851}%
\special{pa 5410 1851}%
\special{fp}%
\special{pa 5635 1770}%
\special{pa 5230 1770}%
\special{fp}%
\special{pa 5518 1689}%
\special{pa 5041 1689}%
\special{fp}%
\special{pa 5212 1608}%
\special{pa 5032 1608}%
\special{fp}%
\special{pa 5635 1932}%
\special{pa 5590 1932}%
\special{fp}%
%
\special{pn 4}%
\special{pa 5554 2337}%
\special{pa 5374 2337}%
\special{fp}%
\special{pa 5626 2256}%
\special{pa 5284 2256}%
\special{fp}%
%
\special{pn 4}%
\special{pa 4933 1689}%
\special{pa 4654 1689}%
\special{fp}%
\special{pa 4852 1608}%
\special{pa 4564 1608}%
\special{fp}%
\special{pa 4672 1527}%
\special{pa 4474 1527}%
\special{fp}%
\special{pa 4492 1446}%
\special{pa 4384 1446}%
\special{fp}%
\special{pa 4843 1770}%
\special{pa 4744 1770}%
\special{fp}%
%
\special{pn 4}%
\special{pa 5032 1608}%
\special{pa 4861 1608}%
\special{fp}%
\special{pa 4897 1527}%
\special{pa 4681 1527}%
\special{fp}%
%
\special{pn 4}%
\special{pa 5635 1527}%
\special{pa 5131 1527}%
\special{fp}%
\special{pa 5635 1446}%
\special{pa 5221 1446}%
\special{fp}%
\special{pa 5635 1365}%
\special{pa 5320 1365}%
\special{fp}%
\special{pa 5635 1284}%
\special{pa 5410 1284}%
\special{fp}%
\special{pa 5635 1203}%
\special{pa 5509 1203}%
\special{fp}%
\special{pa 5635 1122}%
\special{pa 5608 1122}%
\special{fp}%
\special{pa 5635 1608}%
\special{pa 5230 1608}%
\special{fp}%
\special{pa 5635 1689}%
\special{pa 5545 1689}%
\special{fp}%
%
\special{pn 4}%
\special{pa 5914 1932}%
\special{pa 5653 1932}%
\special{fp}%
\special{pa 5986 1851}%
\special{pa 5653 1851}%
\special{fp}%
\special{pa 5833 1770}%
\special{pa 5653 1770}%
\special{fp}%
\special{pa 5842 2013}%
\special{pa 5770 2013}%
\special{fp}%
%
\special{pn 4}%
\special{pa 5770 2094}%
\special{pa 5653 2094}%
\special{fp}%
\special{pa 5698 2175}%
\special{pa 5653 2175}%
\special{fp}%
\special{pa 5761 2013}%
\special{pa 5653 2013}%
\special{fp}%
%
\special{pn 8}%
\special{pa 4141 2418}%
\special{pa 3448 2400}%
\special{ip}%
%
\special{pn 8}%
\special{pa 3493 2139}%
\special{pa 3484 2688}%
\special{ip}%
%
\special{pn 8}%
\special{pa 4006 2337}%
\special{pa 3493 2337}%
\special{fp}%
\special{pa 3520 2256}%
\special{pa 3493 2256}%
\special{fp}%
\special{pa 3745 2256}%
\special{pa 3520 2256}%
\special{fp}%
\put(32.6000,-29.0000){\makebox(0,0)[lb]{Fig4. Phase structure of the Stag Hunt game}}%
\put(35.3000,-31.1000){\makebox(0,0)[lb]{with selfish evolutionary rule}}%
%
\special{pn 8}%
\special{pa 6530 360}%
\special{pa 3100 320}%
\special{ip}%
\end{picture}%